# A Reliable Multipath Routing Protocol Based on Link Stability

Juan Xu ,*Member,IEEE,* Ruofan Wang,*Student Member,*Yan Zhang,*Student Member,*Hongmin Huang,*Student Member*

**Abstract**—Wireless NanoSensor Network (WNSN) is a new type of sensor network with broad application prospects. In view of the limited energy of nanonodes and unstable links in WNSNs, we propose a reliable multi-path routing based on link stability (RMRLS). RMRLS selects the optimal path which perfects best in the link stability evaluation model, and then selects an alternative route by the routing similarity judgment model. RMRLS uses tew paths to cope with changes in the network topology. The simulation shows that the RMRLS protocol has advantages in data packet transmission success rate and average throughput, which can improve the stability and reliability of the network.

**Index Terms**— cooperative communication, routing protocol, SWIPT, WNSNs

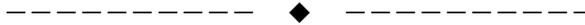

## 1 INTRODUCTION

Due to the development of nanotechnology and the emergence of new materials, the realization and application of Wireless NanoSensor Networks (WNSNs) are feasible. Due to the extremely limited storage capacity of nano-batteries, the communication performance of WNSNs is limited [1,2]. For wireless nano-sensor networks, SWIPT(Simultaneous Wireless Information and Power Transfer) is different from piezoelectric energy harvesting systems. It can provide a stable energy source for nanonodes, so it is a promising charging method.

As an indispensable part of data forwarding and transmission, the routing protocol is also one of the key technologies in WNSNs. Due to the limitation of processing power and storage capacity of nanonodes, traditional routing protocols cannot be directly applied in nano sensor networks. Most WNSNs routing protocols regard reducing energy consumption and extending network lifetime as the primary considerations. In addition, optimizing forwarding paths is also an important research direction in WNSNs. Khan et al. proposed an energy-saving routing protocol for wireless body area networks [3], which divides sensors into two types: critical data sensors and conventional data sensors. For key data, a single-hop method is used for forwarding, and conventional data is transmitted by a combination method of key data sensors single-hop and conventional data sensors multi-hop. This scheme achieves better network stability under the premise of slow death of sensor nodes, and is an efficient routing protocol. Fadi et al. proposed a simple, energy-aware WNSNs routing protocol LaGOON [4]. The LaGOON protocol improves efficiency through some small improvements. For example, the neighbor selected by the source node is included in the message to optimize the forwarding path. The agreement takes into account the limitations of nanonodes and focuses on the two purposes of reducing communication overhead and energy saving. Verma et al. proposed a new energy-saving routing framework based on WNSNs. The framework is designed based on clustering, and solves the energy efficiency problem of WNSN through energy balance clustering. Simulation shows that the routing framework can effectively reduce energy consumption [5]. Although this protocol optimizes the selection process of cluster heads, it cannot solve the problem of cluster heads that are too low in energy to work. Considering that there may be ultra-dense characteristics in nano-sensor networks, Aliouat et al. proposed a novel, efficient, and intuitive distributed routing protocol for ultra-dense terahertz networks (M2MRP). The protocol uses node density to design a robust routing strategy and moderate additional flow control. The simulation results show that M2MRP is superior to the known dense network routing protocol in terms of the number of information exchanges and the success rate [6]. Although the above routing protocols have carried out some researches on reducing energy consumption or routing strategies, they have not combined energy harvesting methods to fundamentally solve the energy consumption problem. The energy consumption problem is still an unbreakable bottleneck in WNSNs.

For the WNSNs link instability and the poor adaptability of the energy balance cluster network framework (EBCNF) to network topology changes, a reliable multi-path routing based on link stability (RMRLS) is proposed. The link stability is evaluated, the optimal path is selected based on the link stability, and then a routing similarity judgment

————————————————

- *J. Xu is with the College of Electronics and Information Engineering, Tongji University, Shanghai, 201804, China (e-mail: jxujuan@tongji.edu.cn).*
- *R. Wang is with the College of Electronics and Information Engineering, Tongji University, Shanghai, 201804, China(e-mail: 1047990986@qq.com).*
- *Y. Zhang is with the College of Electronics and Information Engineering, Tongji University, Shanghai, 201804, China(e-mail: 1830732@tongji.edu.cn).*
- *H. Huang is with the College of Electronics and Information Engineering, Tongji University, Shanghai, 201804, China (e- mail: freeastime@163.com).*

model is proposed. Through this model, another route with the lowest similarity to the optimal route is selected as alternative route, and finally two routing paths, the main path and the backup path, are established to cope with changes in the network topology and improve the stability of data transmission.

The remainder of this paper is organized as follows. In Section 2, the system model is introduced. In Section 3, inter-cluster routing protocol RMRLS is described. And then, we demonstrate and analyze simulation results for the performance of RMRLS protocol in Section 4. Finally, we conclude this paper in Section 5.

## 2 SYSTEM MODEL

### 2.1 Route Similarity Judgment Model

The RMRLS protocol proposed in this paper adopts the way of replacing multipath for data transmission. Replacing multipath refers to separating multiple routing paths from primary to secondary. Only when the primary path fails, the backup path is used to replace the primary path for data transmission.

On the basis of the known main path, the relationship between multiple paths and the main path is quantified by designing a route similarity judgment model, and the alternate route is selected. The specific routing similarity calculation method is as follows:

$$\Omega_i = k(N-2) + \sigma L \qquad (1)$$

where $N$ indicates that the current path $p_i$ and the main path have $N$ nodes that are intersected, $L$ indicates that the current path $p_i$ and the main path have $L$ links that are shared. $k$ and $\sigma$ are the weights of node correlation and link correlation, and $0 < k, \sigma < 1$ and $k + \sigma = 1$.

Figure 1 is an example to illustrate the calculation process of similarity between routes. Taking A-B-D-G-H-J in Figure 1 as the main route and A-C-E-F-I-G-H-J as a candidate route, then the two routes have 3 nodes that are intersected, and the two links, G-H and H-J, are shared. Assuming that $k$ and $\sigma$ are both 0.5, the routing similarity $\Omega$ of the two paths is 1.5.

Based on the known main path, the routing similarity between multiple candidate paths and the main path is calculated and sorted, and the path with the lowest similarity is selected as the backup path. If there is more than one path with the lowest correlation, the path with better quality is selected as the candidate path using link stability estimation.

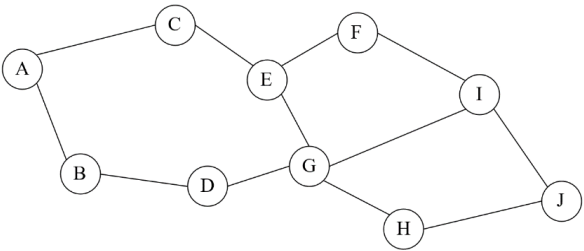

Fig. 1. Schematic diagram of the routing similarity

### 2.2 Link State Estimation Model

In order to estimate the link state better, we use Kalman filter [7] to estimate the link state. The Kalman filter can efficiently integrate all kinds of information and make predictions. Different from traditional prediction algorithms, Kalman filtering can effectively use the information in historical data, and the more data, the better its performance. After the path loss between nodes is obtained, the received power between nodes can be calculated as:

$$P_r = \frac{P_t}{PL} \qquad (2)$$

where $P_r$ is the received power, $P_t$ is the transmit power, and $PL$ is the path loss.

When the first set of $P_r$ between nodes is calculated, the Kalman filter takes it as its input, and then uses equations (3)-(7) to iteratively find the prior value of the received power and the parameter covariance matrix ($O$). The process equation is as follows:

$$P_r^-(t_n) = K \cdot P_r^+(t_{n-1}) + n(t_n) \qquad (3)$$

where $P_r^-(t_n)$ is the estimated received power at time $t_n$, and $K$ is the state transition parameter of the linear system. Since there is only one parameter for the current link state estimation model, $K$ is a constant. $P_r^+(t_{n-1})$ is the best predicted power at time $t_{n-1}$ after the correction. $n(t_n)$ is the process noise from $t_{n-1}$ to $t_n$. We assume that the noise is Gaussian white noise and satisfies $n(t_n) \sim (0, Q)$.

The update process of the parameter covariance matrix can be expressed as:

$$O^-(t_n) = KO^+(t_{n-1})K^T + Q \qquad (4)$$

where $Q$ is the covariance of the process noise $n(t_n)$, is the parameter covariance matrix at $t_n$, and $O^+(t_{n-1})$ is the corrected parameter covariance matrix at $t_{n-1}$.

In addition to prediction errors, sensors also have measurement errors. We use the signal strength received by the nano sensor to indicate RSSI as the measured value of the received power. Suppose the measured value of the sensor is $x(t_n)$, and the measurement noise is represented by $w(t_n)$. We assume that the noise is Gaussian white noise and satisfies $w(t_n) \sim (0, Z)$.

Then, the Kalman gain can be expressed as:

$$M(t_n) = O^-(t_n)H^T[HO^-(t_n)H^T + Z]^{-1} \qquad (5)$$

where $Z$ is the covariance of the measurement noise, and $H$ is the measurement system parameter.

The update equation of the received power is:

$$P_r^+(t_n) = P_r^-(t_n) + M(t_n)[x(t_n) - H \cdot P_r^-(t_n)] \qquad (6)$$

The update equation of the parameter covariance matrix is:

$$O^+(t_n) = [1 - M(t_n)H]O^-(t_n) \qquad (7)$$

Finally, the link quality between node $i$ and node $j$ in time slot $t_n$ can be calculated as follows:

$$L^{(i,j)+}(t_n) = \frac{1}{1 + e^{-\left(P_r^{(i,j)+}(t_n) - \vartheta\right)}} \qquad (8)$$

where $\vartheta$ is the standard deviation of the first group of received power of the Kalman filter. In this way, the end-to-end link quality of the $t_n$ time slot can be predicted as a value between 0 and 1.

## 3 INTER-CLUSTER ROUTING PROTOCOL RMRLS

Based on the EBCNF framework, to address the problem of low reliability of routing data transmission between clusters on a single path, we propose a reliable multi-path routing based on link stability (RMRLS). The RMRLS protocol is an on-demand routing. The routing is only established when the nodes in the network need to transmit data. Compared with the preemptive routing, this protocol has less routing overhead and is more suitable for sensor networks.

The RMRLS protocol includes two stages: route establishment and route maintenance. In the route establishment phase, the node first discovers the effective neighbor nodes around it through simple Hello messages, and then evaluates the link stability of each candidate node. The first $m$ nodes with higher stability are used as next hop nodes. The route maintenance stage is mainly to deal with the situation of forwarding failure and transmission interruption in the process of data forwarding.

### 3.1 Candidate Node Discovery Mechanism

The RMRLS protocol introduces a candidate node discovery mechanism, through which the energy state of neighbor nodes is judged. Only when the neighbor node has enough energy to perform the route establishment process will it be added to the candidate node set for subsequent processes.

In the RMRLS protocol, nodes request and respond to messages through the following datagrams: neighbor discovery messages (NDIS), neighbor feedback messages (NFEE), route request messages (RREQ), and route response messages (RREP). The number of bits of these messages are represented by $N_{NDIS}$、$N_{NFEE}$、$N_{PREQ}$ and $N_{PREP}$ respectively. The specific steps of the candidate node discovery mechanism are as follows:

(1) The nanonode $u_x$ broadcasts a neighbor discovery message NDIS to neighbor nodes within the communication range;

(2) After receiving the NDIS message, the neighbor node $u_y$ compares whether its remaining energy $E_y$ is greater than the energy threshold, which can be expressed as:

$$E_y \geq \epsilon(N_{NDIS}E_{bit} + N_{NFEE}E_{bit} + N_{PREQ}E_{bit} + N_{PREP}E_{bit}) \qquad (9)$$

where $\epsilon$ is the system constant, and the energy threshold is the energy required for the node to send and receive several messages. If the inequality is satisfied, the neighbor node $u_y$ replies with a neighbor feedback message, which contains the neighbor node's own ID, the remaining energy value, and the ID of the nanonode $u_x$. If it is not satisfied, the neighbor node $u_y$ does not respond to this NDIS message.

### 3.2 Path Selection Based on Link Stability Evaluation

In many protocols, such as AOMDV and OLSR, the shortest path is selected as the routing path. However, if the nodes on these paths have insufficient energy or poor link status, data transmission may be interrupted. Therefore, when selecting a routing path, we comprehensively evaluate the remaining energy, link quality and distance of the nodes on the link, and select the next hop node based on this. The three factors of remaining energy, link quality, and distance are all reported to the current node by the candidate node through neighbor feedback messages. First, the candidate node reports its remaining energy to the source node through neighbor feedback messages. The higher the remaining energy, the node can stably transmit more information; the link quality is estimated by Kalman filter, the higher the link quality, the node is more suitable as the next hop node; the distance is the distance between the candidate node and NC. Under the premise of ensuring the link is reachable, the smaller the distance between the candidate node and the NC, then the transmission path require smaller hops to transmit data.

When measuring the influence of node's remaining energy, link quality and distance on link stability, we use entropy method to analyze. In the entropy method, referring to the viewpoint of information theory, it believes that the larger the entropy, the smaller the information known, the greater the uncertainty of the factor, and the smaller the utility value, the smaller the weight. This view is also applicable to link stability assessment. For links, the greater the entropy of this factor, the greater the uncertainty of the factor, the worse the stability, and the smaller the weight.

Assuming that the source node obtains the remaining energy, link quality, and distance of $n$ candidate nodes, then the initial data matrix for link stability evaluation can be obtained:

$$B = \begin{bmatrix} b_{11} & b_{12} & b_{13} \\ \vdots & \vdots & \vdots \\ b_{n1} & b_{n2} & b_{n3} \end{bmatrix}_{n \times 3} \qquad (10)$$

where $b_{ij}$ represents the data of the $j$-th factor of the $i$-th node.

After obtaining the original data, the data needs to be standardized. In addition, in the entropy method, there is a difference between positive index data and negative index data. Positive index data indicates that the higher the value of this type of data, the better, and the negative index data indicates that the lower the value of this type of data, the better. For these two types of data, different methods need to be used for data standardization.

For positive indicator data, the processing method is:

$$b'_{ij} = \frac{b_{ij} - min\{b_{1j}, \dots, b_{nj}\}}{max\{b_{1j}, \dots, b_{nj}\} - min\{b_{1j}, \dots, b_{nj}\}} \qquad (11)$$

For negative indicator data, the processing method is:

$$b'_{ij} = \frac{max\{b_{1j}, \dots, b_{nj}\} - b_{ij}}{max\{b_{1j}, \dots, b_{nj}\} - min\{b_{1j}, \dots, b_{nj}\}} \qquad (12)$$

According to the definition of positive index data and negative index data, it can be concluded that the remaining energy and link quality data belong to the positive index data, which needs to be processed by equation (11). And the distance is negative, so it needs to be processed by equation (12).

After the standardization process, it is necessary to calculate the proportion of each data in the evaluation factor:

$$p_{ij} = \frac{{b_{ij}}'}{\sum_{i=1}^{n} {b_{ij}}'} \qquad (13)$$

After obtaining the proportion of each data, the entropy value of each evaluation factor can be calculated according to the entropy calculation formula:
$$e_j = -k \sum_{i=1}^{n} p_{ij} \ln(p_{ij}) \quad (14)$$
where $k$ is a constant, and its value depends on the number of current candidate nodes $n$. It can be calculated by the following equation:
$$k = \frac{1}{\ln(n)} \quad (15)$$

After obtaining the entropy value of each factor, it is necessary to calculate the information utility value of each factor. The information utility value of a factor depends on the difference between the information entropy of the factor and 1. The greater the information utility value, the greater the importance of this factor to the stability of the final link. The information utility value is calculated by the following equation:
$$z_j = 1 - e_j \quad (16)$$

After obtaining the information utility value of each factor, the weight of each factor can be calculated by the following equation:
$$w_j = \frac{z_j}{\sum_{j=1}^{3} z_j} \quad (17)$$

After obtaining the weight of each factor, the link stability between the source node $u_s$ and each candidate node $u_i$ can be calculated:
$$s(u_s, u_i) = \sum_{j=1}^{3} w_j \, b_{ij}' \quad (18)$$

The higher the calculated link stability, the better the link stability between the candidate node and the source node, and the more likely the candidate node is to be selected as the next hop node. Due to the limited energy of nanonodes in WNSNs, if all candidate nodes perform data forwarding to find routing paths, the burden on the network is too large, so we use link stability to evaluate the links of $n$ candidate nodes and select the $m$ highest stability nodes as next hop nodes, and the specific calculation method of the number of next hop nodes can be expressed as:
$$m = \begin{cases} n & n \leq \tau \\ \lfloor n/\tau \rfloor & n > \tau \end{cases} \quad (19)$$
where $\tau$ is a system parameter. By processing the number of next-hop nodes with the above equation, the calculation amount of nanonodes when generating the forwarding list can be reduced, and the scope of routing flooding in the network can be further reduced, and the burden on the network can be reduced.

### 3.3 Route Establishment Phase

The process of establishing the RMRLS routing algorithm is shown in Figure 6. The specific steps are as follows:

(1) RMRLS routing is an on-demand routing. When a cluster head node $u_s$ needs to send the sensed data to NC, the routing will be established. First, the source cluster head node $u_s$ checks its local routing cache to find a path to NC. If there is a path to NC in the local routing cache, data transmission is performed according to this path; if there is no path to NC, the candidate node discovery mechanism is executed, and if there is NC in the set of candidate nodes, then the data is directly forwarded to NC; Otherwise, go to step (2).

(2) The source cluster head node $u_s$ evaluates the link stability of each candidate node, and selects the first $m$ candidate nodes with the highest link stability as the next hop nodes. The source cluster head node $u_s$ sends a routing request message (RREQ) containing the link stability of the current next hop node and current routing information to the corresponding next hop node. The route request message (RREQ) contains the source node ID, the destination node ID (ie the ID of NC), the routing record (cumulatively record the ID of the nanonodes that this routing request message passes through in order), and the request ID (sequence number generated by the source node) and the total link stability of the current routing record (the link stability in the previous RREQ plus the current link stability).

(3) For the intermediate node $u_c$, after receiving the route request message, it needs to detect the received RREQ message. If it is detected that the received RREQ is unprocessed, the intermediate node $u_c$ adds its ID to the routing record, and responds with an ACK message to the previous hop node, and then performs the same steps as the source node $u_s$, namely step (1) (2); if the route cache of the intermediate node $u_c$ contains the path to NC, the intermediate node $u_c$ adds its ID to the routing record and adds the link stability of the next hop node in the routing cache to the total link stability, the RREQ message is then directly sent to the next hop node in the routing buffer, and an ACK message is returned to the previous hop node; if the routing record in the RREQ message received by the intermediate node $u_c$ already contains its own ID or has received the same RREQ message, the request message is discarded.

(4) Within the specified time, the NC collects as many paths information as possible, and according to the total link stability in the final RREQ message, selects the path with the highest medium stability as the main path, and the path with the lowest similarity to the main path as a backup path to prevent network topology changes, and sends a routing response message (RREP) to the source node $u_s$ to indicate that the path is established. After receiving the RREP message, the source node $u_s$ updates its routing table information.

### 3.4 Route Maintenance Phase

In the RMRLS protocol, the source node preferentially selects the main path for data transmission. Since the nanonode in WNSNs can perceive the connection status of the next hop node through simple Hello messages, once it finds that there is a link interruption or data forwarding failure during the data transmission process, the current node will immediately report the error information to the source node. The node deletes the routing information currently in error from the routing table, and switches to a backup path for data transmission.

## 4 SIMULATION ANALYSIS

### 4.1 Simulation Parameter

The simulation scene is a two-dimensional square plane of 10mm×10mm. Nanonodes are randomly distributed in this plane. There is one and only one NC located on the right side of the square area, which is balanced with the center line of the square area, and is 1mm from the right side of the square (11mm, 5mm), as shown in Figure 2.

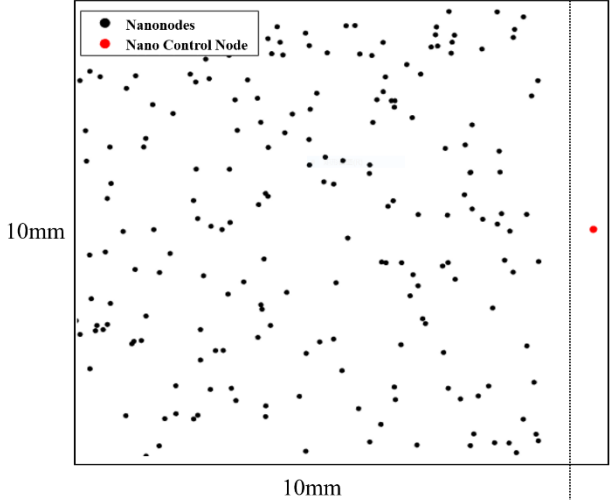

Fig. 2. Network nodes deployment

The constant $\epsilon$ in equation (9) is 1. The remaining energy threshold $i$s set to $1.4 \times 10^{-13}$ J. The system parameter $\tau$ in equation (19) is set to 2. The lengths of various datagrams in the RMRLS protocol are: $N_{NDIS} = N_{NFEE} = 2$Byte、$N_{PREQ} = N_{PREP} = 6$Byte.

In the simulation, the RMRLS protocol establishes two feasible paths in a route discovery process. Taking NC as the destination node, each time 20 nodes are randomly distributed in different distance ranges (the straight-line distance between the node and NC) to generate data packets for data transmission. In the simulation results, the different distance ranges are simplified to numbers, such as 1 means within the range of 0-1, 2 means within the range of 1-2. The specific simulation parameters are shown in Table 1.

TABLE 1
SIMULATION PARAMETER

| Simulation parameters (unit) | Numerical value |
|---|---|
| Simulation scene (mm²) | 10×10 |
| Number of Nano Control Nodes | 1 |
| Number of nano nodes | 200 |
| Number of iterations | 1500 |
| Data packet length (Byte) | 128 |
| Distance between source node and destination node (mm) | [1:1:10] |
| Signal propagation speed $v$ (m/s) | $3 \times 10^8$ |
| Data packet generation interval (s) | 0.01 |
| WET time slot duration (s) | 5 |
| SWIPT time slot duration (s) | 0.01 |
| WIT time slot duration (s) | 0.1 |
| Data packet generation interval (s) | 0.001~0.01 |
| Initial energy of nano node ($\mu$J) | 4 |
| Simulation time (s) | 120 |

## 4.4 RMRLS Protocol Simulation

From the perspective of receiving architecture, the structure of TS mechanism is simpler in practice and easier to implement. Therefore, TS-EBCNF framework is adopted for comparison here.

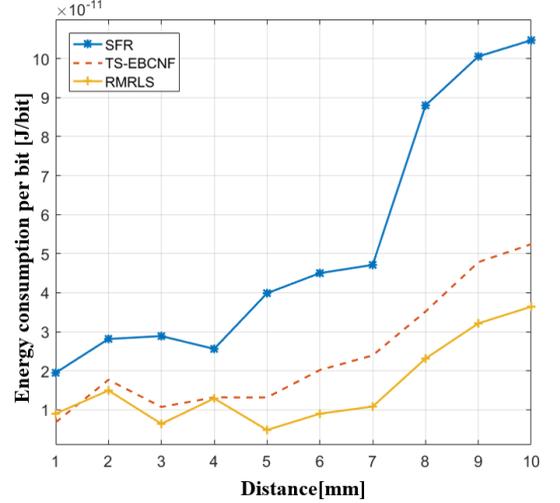

Fig. 3. Comparison of energy consumption per unit bit

(1) Energy consumption per unit bit

Due to the existence of the SWIPT mechanism, the nanonodes in the network have energy replenishment, so it is impossible to directly judge the energy consumption of the network from the remaining energy value of the network. We use the energy consumption per bit to more intuitively judge the impact of the three protocols on the energy consumption of the network. Figure 3 compares the energy consumption per unit bit of the three protocols. As the distance between the source node and NC increases, the energy consumption per unit bit of the three protocols gradually increases, and the increase in energy consumption per unit bit of the SFR protocol is significantly greater than that of the other two protocols. This is because the SFR protocol is a routing protocol based on selective flooding. When the distance increases and the data transmission between the source node and the destination node requires more hops, more nanonodes in the network will participate in routing flooding process, so more energy will be wasted for data transmission. RMRLS considers the impact of link stability on data transmission on the basis of TS-EBCNF, and establishes a more stable link for data transmission. Compared with the SFR protocol, the RMRLS protocol adopts a multipath alternative. The scheme is used for data transmission, and the backup path is activated only when there is a problem with the main path. Compared with the SFR protocol using multiple paths for data transmission through flooding, it can better reduce the energy consumption of the nanonodes.

(2) Data packet transmission success rate

Figure 4 is a comparison of the data packet transmission success rate of the three protocols. When the distance between the source node and NC is less than 2mm, the data packet can be forwarded to the NC 100%, because the source node only needs one hop to directly forward the data packet to the NC. As the distance between the source node and NC increases, the success rate of data packets decreases significantly, and when the distance is greater than 6mm, the transmission success rate of the RMRLS protocol is significantly better than the other two protocols. This is because the RMRLS protocol considers the influence of residual energy, link quality, and distance on link stability during the process of establishing a route, which improves the reliability of the network. Since TS-EBCNF only randomly selects the next hop node for data forwarding, it does not consider the link status between the current node and the next hop node. However, when the distance is far, the energy consumption of nodes in the network using the SFR protocol increases, and the molecular absorption loss between the nodes is serious, and the channel quality is poor, resulting in a significant decrease of data packet transmission success rate.

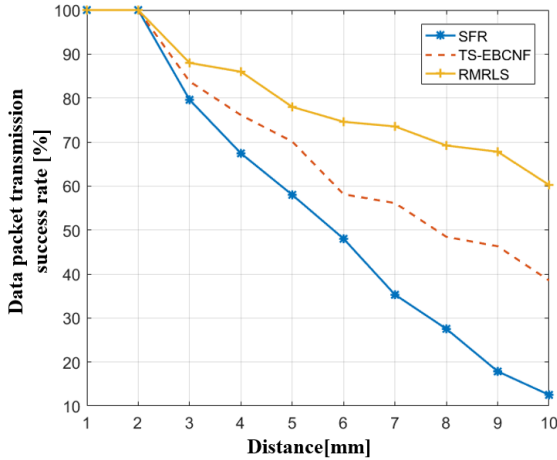

Fig. 4. Comparison of data packet transmission success rate

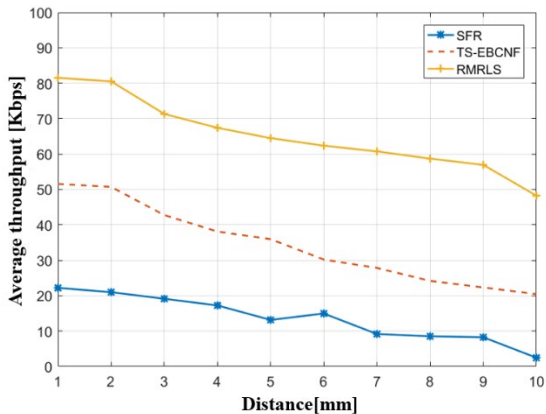

Fig. 5. Comparison of average throughput

(3) Average throughput

Figure 5 is a comparison of the average throughput of the three protocols. The RMRLS protocol can better overcome the impact of distance on throughput and can maintain a high throughput in the case of a small distance. This is because first, from the previous analysis, it can be seen that RMRLS has higher data packet transmission success rate than the other two protocols; second, compared with the SFR protocol, RMRLS has higher energy efficiency which avoids the inability to transmit data due to the high energy consumption of nano nodes in the SFR protocol.

## 5 CONCLUSION

In view of the fact that the energy of nanonodes in WNSNs is exhausted and cannot work or the nodes are damaged due to environmental changes, as well as the shortcomings in the EBCNF framework, we improve the inter-cluster routing in the EBCNF framework and propose a reliable multi-path routing protocol RMRLS. This protocol is an on-demand multi-path routing protocol. Through the evaluation of the link stability between nodes, the path with the highest link stability is selected for data transmission to improve the stability of data transmission, and secondly through routing similarity judge model and selecting alternative route, in order to deal with the situation that the nanonodes in the network may run out of energy and cannot work or the environment changes cause damage to the nodes, and improve the reliability of the network. The final simulation results verify that the RMRLS protocol has advantages in data packet transmission success rate and average throughput, and can improve the stability and reliability of the network.